\documentclass[aps,preprint,showpacs,preprintnumbers,amsmath,amssymb]{revtex4}
\usepackage[dvips]{graphicx}

\begin{document}

\title{Fermion production in Coulomb field on de Sitter universe}
\author{Crucean Cosmin}
 \email{crucean@quasar.physics.uvt.ro}
 \affiliation{Faculty of Physics, West University of Timi\c soara,  \\V. Parvan
 Avenue 4 RO-300223 Timi\c soara,  Romania}

\date{\today}

\begin{abstract}
Fermion production in an external Coulomb field on de Sitter expanding universe is studied. The amplitude and probability of pair production in an external Coulomb field are computed and the cases of large/small values of the expansion factor comparatively with the particle mass are studied. We obtain from our calculations that the modulus of the momentum is no longer a conserved quantity. We find that in the de Sitter space there are probabilities for production processes where the helicity is no longer conserved. For a vanishing expansion factor we recover the Minkowski limit where the amplitude of this process vanishes. The rate of pair production in Coulomb field is found to be important in the early universe when the expansion factor was large comparatively with the particle mass.
\end{abstract}


\maketitle

\section{Introduction}
The problem of pair production in an external gravitational field was intensively studied in the literature beginning with the Schr\"{o}dinger paper \cite{1}, who first suggested that in an expanding universe particle production could arise. The next important step in this direction was done by L.Parker \cite{7,8,9} who understood quantitatively the phenomenon of pair production in a gravitational field. The main results of L.Parker's work are related to the computation of the number density of scalar particles and Dirac particles created in an expanding background. Also, the limit of zero mass and infinite mass was addressed and in addition, an estimation of the amount of particle production for the present day expansion was made. Therefore in the present paper we will compare our results with these important predictions made by L.Parker in \cite{7,8,9}, even if we will use another approach based on perturbations.

In this paper we will follow the usual quantum field theory approach, were the transition amplitudes are defined as in \cite{12} using perturbation methods. This approach was also used in \cite{15,16,17},\cite{20,21,22}, where transitions that generate particle production from vacuum were also studied on Friedmann-Robertson-Walker metrics.
With this general approach, we can study the limit of these transition amplitudes calculated on de Sitter QED, for a vanishing expansion factor. This corresponds to the flat space limit, where the amplitude of this process vanishes and the connection with the Minkowski QED can be done. Also, from our general result, one could address the interesting case of large expansion, which corresponds to the early universe where the rate of pair production becomes important. As it was well established in the literature, particle production could arise as the result of the fact that the in and out vacua are not the same \cite{18,19,27}, or because of the loss of the translational invariance with respect to time \cite{12} which in loose terms refers to the well known fact that in this geometry the laws of conservation for energy and momentum are lost\cite{7,8,9,12}. It is well known that in the Minkowski theory \cite{5,6,13,14} the pair production in a Coulomb field could not arise as a perturbative phenomenon. Contrary to this, in de Sitter space the translational invariance with respect to time is lost and one should expect a nonvanishing probability of transition for the process: $vacuum \rightarrow e^-+e^+$ in the presence of a Coulomb potential. It is also important to mention that the electron-positron pair could also be absorbed by the Coulomb field in the de Sitter geometry, which represents the time-reversed process. The transition amplitudes will be calculated here using only the mode expansion in momentum basis, which have a defined momentum and helicity. The fermion production in a Coulomb field in de Sitter space was to the best of our knowledge not studied so far. We also specify that our results are in good agrement with the work previously done on the particle production in a pure gravitational field, which predicts important amounts of particle creation in the early stage of the universe \cite{7,8,9}.

It is worth to mention that important results were obtained in \cite{18,19}, where the production of fermionic particles and scalar particles in electric fields on de Sitter geometry was considered, using a nonperturbative approach. With this approach one can obtain the density number of particles created in the far out region and because there are some constrains related to the mass of the particle \cite{19,27}, the interesting case of large expansion from the early universe is not properly addressed. For the Dirac field, the results presented in \cite{18}  show that there are nonvanishing rates for fermion production of mass zero and of finite mass in electric fields in de Sitter space. A few comments need to be made here related to the production of zero mass fermions. In \cite{7,8,9} and more recently in \cite{27} it was proven that because in the massless case the field is conformally coupled, there is no production of fermions with zero mass. It will be interesting to compare our result which was obtained using the exact solutions of the field equation with the above mentioned results\cite{7,8,9,18,19,27}, in the massless case. This is because the results from \cite{7,8,9,27}, seems to be in contradiction with \cite{18,19}, in the null mass case.

From another point of view, it is important to directly prove that the transition amplitudes of pair creation in an external field contain terms that break the conservation laws, which seems to have been known for a long time but was never proven in a direct calculation using the solutions of free field equations. For these reasons, we hope that the results presented here will be of interest for cosmology and quantum field theory on Friedmann-Robertson-Walker backgrounds.

The second section of our paper begins with a short reminder of the work that was done in the theory of free fields on de Sitter QED. In the third section we present the main steps for calculating the transition amplitude corresponding to the process of pair production in the Coulomb potential in de Sitter geometry and we compute the probability for pair production. Physical consequences of our calculations are also analyzed in this section. The fourth section is dedicated to the study of limit cases as the expansion factor is small/large comparatively with the mass of the particle. Our conclusions are summarized in section five.

\section{Free fields on de Sitter space}

The definition of transition amplitudes is expressed with the help of the exact solutions
of the free Dirac and Maxwell equations on the de Sitter geometry, expanded in the momentum basis. For this reason, we make a short review of the results from free field theory in this background. The line element \cite{2} for de Sitter universe is:
\begin{equation}\label{metr}
ds^{2}=dt^{2}-e^{2\omega t}d\vec{x}^{2}
\end{equation}
where $\omega$ is the expansion factor and $\omega>0$.  Now we know that for defining fields with half integer spin on curved spacetime it is required to use the tetrad fields \cite{2} $e_{\widehat{\mu}}(x)$ and $\widehat{e}^{\widehat{\mu}}(x)$,
fixing the local frames and corresponding coframes which are labelled by the local
indices $\widehat{\mu},\widehat{\nu},...=0,1,2,3$. The form of the
line element allows us to choose the simple Cartesian gauge with
the nonvanishing tetrad components
\begin{equation}
e^{0}_{\widehat{0}}=e^{-\omega t}  ;e^{i}_{\widehat{j}}=\delta^{i}_{\widehat{j}}\,e^{-\omega t}
\end{equation}
so that $e_{\widehat{\mu}}=e^{\nu}_{\widehat{\mu}}e_{\nu}$ and
have the orthonormalization properties
$e_{\widehat{\mu}}e_{\widehat{\nu}}=\eta_{\widehat{\mu}\widehat{\nu}},\\
  \widehat{e}^{\widehat{\mu}}e_{\widehat{\nu}}=\delta^{\widehat{\mu}}_{\widehat{\nu}}$
with respect to the Minkowski metric $\eta=diag(1,-1,-1,-1)$.

Now let us introduce unit normalized helicity spinors \cite{6} for an
arbitrary momentum vector $\vec{p}$ and denote them
$\xi_{\lambda}(\vec{p}\,)$ and $\eta_{\sigma}(\vec{p}\,)= i\sigma_2
[\xi_{\sigma}(\vec{p}\,)]^{*}$
\begin{equation}\label{pa}
\vec{\sigma}\vec{p}\,\xi_{\lambda}(\vec{p}\,)=2p\lambda\xi_{\lambda}(\vec{p}\,)
\end{equation}
with $\lambda=\pm1/2$ and where $\vec{\sigma}$ are the Pauli
matrices and $p=\mid\vec{p}\mid$.
The particle spinors have the form
\begin{equation}\label{xi}
\xi_{\frac{1}{2}}(\vec{p}\,)=\sqrt{\frac{p_3+p}{2 p}}\left(
\begin{array}{c}
1\\
\frac{p_1+ip_2}{p_3+p}
\end{array} \right)\,,\quad
\xi_{-\frac{1}{2}}(\vec{p}\,)=\sqrt{\frac{p_3+p}{2 p}}\left(
\begin{array}{c}
\frac{-p_1+ip_2}{p_3+p}\\
1
\end{array} \right)\,.
\end{equation}

Then, the positive/negative frequency modes of momentum $\vec{p}$ and
helicity $\lambda$ that were derived in \cite{3}, assuming that gamma
matrices in Dirac representation, are:
\begin{eqnarray}\label{sol}
U_{\vec{p},\lambda}(t,\vec{x}\,)=\frac{\sqrt{\pi
p/\omega}}{(2\pi)^{3/2}}\left (\begin{array}{c}
\frac{1}{2}e^{\pi k/2}H^{(1)}_{\nu_{-}}(\frac{p}{\omega} e^{-\omega t})\xi_{\lambda}(\vec{p}\,)\\
\lambda e^{-\pi k/2}H^{(1)}_{\nu_{+}}(\frac{p}{\omega} e^{-\omega
t})\xi_{\lambda}(\vec{p}\,)
\end{array}\right)e^{i\vec{p}\cdot\vec{x}-2\omega t};\nonumber\\
V_{\vec{p},\lambda}(t,\vec{x}\,)=\frac{\sqrt{\pi
p/\omega}}{(2\pi)^{3/2}} \left(
\begin{array}{c}
-\lambda\,e^{-\pi k/2}H^{(2)}_{\nu_{-}}(\frac{p}{\omega} e^{-\omega t})\,
\eta_{\lambda}(\vec{p}\,)\\
\frac{1}{2}\,e^{\pi k/2}H^{(2)}_{\nu_{+}}(\frac{p}{\omega} e^{-\omega t}) \,\eta_{\lambda}(\vec{p}\,)
\end{array}\right)
e^{-i\vec{p}\cdot\vec{x}-2\omega t},
\end{eqnarray}
where $H^{(1)}_{\nu}(z), H^{(2)}_{\nu}(z)$ are Hankel functions of the first and second kind and $k=\frac{m}{\omega},\nu_{\pm}=\frac{1}{2}\pm ik$. It is also worth to mention that these solutions for the Dirac field in the Robertson-Walker metric were obtained earlier in \cite{4}, with the specification that the normalization constants were not determined. As was shown in \cite{3}, these constants depend on physical quantities such as momentum and play a central role in an exact calculation. So  we will use in our further considerations the solutions from \cite{3}.

These solutions are normalized such that \cite{3}:
\begin{equation}
\int d^{3}x
(-g)^{1/2}\bar{U}_{\vec{p},\lambda}(x)\gamma^{0}U_{\vec{p^{\prime}},\lambda^{\prime}}(x)=
\delta_{\lambda\lambda^{\prime}}\delta^{3}(\vec{p}-\vec{p^{\prime}}),
\end{equation}
with the specification that the $V$ solutions obey similar relations and the two modes are orthogonal to each other.
For a better explanation of what mean positive/negative frequencies we can use the expansion of the Hankel functions:
\begin{equation}
H^{(1,\,2)}_{\nu}(z)\simeq\left(\frac{2}{\pi z}\right)^{1/2}e^{\pm i(z-\nu\pi/2-\pi/4)}.
\end{equation}
For $t\rightarrow-\infty$, the modes defined above behave as a positive/negative frequency modes with respect to conformal time $t_{c}=-e^{-\omega t}/\omega$:
\begin{equation}
U_{\vec{p},\lambda}(t,\vec{x}\,)\sim e^{-ipt_{c}};\,\,V_{\vec{p},\lambda}(t,\vec{x}\,)\sim e^{ipt_{c}},
\end{equation}\label{bd}
which is precisely the behavior in the infinite past for positive/negative frequency modes which defines the Bunch-Davies vacuum \cite{12}.
\par
As in Minkowski space, the negative frequency modes are obtained with the charge conjugation operation \cite{5,13}
\begin{equation}\label{charge}
U_{\vec{p},\lambda}(x)\rightarrow
V_{\vec{p},\lambda}(x)=i\gamma^{2}\gamma^{0}(\bar{U}_{\vec{p},\lambda}(x))^{T},
\end{equation}
 because the charge conjugation symmetry remains valid in a curved background.
\par
In our calculation, we need the form of the Coulomb potential on de
Sitter spacetime which is dependent on the line element. The situation here is simple if we recall the conformal invariance of Maxwell's equations \cite{28}. The Coulomb field in Minkowski space is $A^{0}=\frac{Ze}{|\vec{x}|}$. Then one finds for
the corresponding de Sitter potential the following formula:
\begin{equation}\label{pot}
A^{\widehat{0}}(x)=\frac{Ze}{|\vec{x}|} e^{-\omega
t},A^{\widehat{j}}(x)=0,
\end{equation}
where the hatted indices indicate that we refer to the components
in the local Minkowski frames $A^{\widehat{\mu}}=e^{\widehat{\mu}}_{\nu}A^{\nu}$. We also observe that (9) is just
the expression from flat space with spatial distances dilated/contracted
by a factor $e^{-\omega t}$.

\section{Amplitude calculation}
In this section we will calculate the transition amplitude for pair production in a Coulomb field and we will explore the physical consequences of our result.
The form of the transition amplitude can be established using the same methods as in the flat space case, as was shown in \cite{12,15,20,21}. For pair production in an external electromagnetic field, supposing that the fields are coupled by the elementary electric charge $e$, the expression of the transition amplitude is:
\begin{equation}\label{ampl}
\mathcal{A}_{e^-e^+}=-ie \int d^{4}x
\left[-g(x)\right]^{1/2}\bar U_{\vec{p},\,\lambda}(x)\gamma_{\mu}A^{\widehat{\mu}}(x) V_{\vec{p}\,\,',\,\lambda'}(x)
\end{equation}
where $e$ is the unit charge of the field and the fields $U_{\vec{p},\,\lambda}(x)\,\,,V_{\vec{p}\,\,',\,\lambda'}(x)$ are supposed to be exact solutions of the free Dirac equation in the momentum basis.
The amplitude defined above for particle production in the potential (\ref{pot}), can now be expressed with the help of the solutions of free Dirac equation (\ref{sol}). The integral that results splits in a spatial integral and a temporal integral. The form of the spatial integral is better known from flat space theory \cite{6} and this result is incorporated in the next formula for the amplitude:
\begin{eqnarray}\label{in}
\mathcal{A}_{e^-e^+}=i
\frac{e^{2} Z\sqrt{pp\,'}}{8\pi|\vec{p}+\vec{p}\,'|^{2}}\left[-sgn(\lambda\,')\int_0^{\infty} dz
zH^{(2)}_{\nu_{+}}(p z)H^{(2)}_{\nu_{-}}(p\,'z)\right.\nonumber\\
\left.+sgn(\lambda)\int_0^{\infty} dz
zH^{(2)}_{\nu_{-}}(p z)H^{(2)}_{\nu_{+}}(p\,' z)\right]\xi^{+}_{\lambda}(\vec{p}\,)\eta_{\lambda'}(\vec{p}\,\,'),
\end{eqnarray}
where the new variable of integration is $z=-t_{c}$ and $sgn$ is the signum function. The temporal integral which contains the influence of the gravity in this process is not so simple. We shall give here only the main steps for solving the integrals.
Using the relations between Hankel and Bessel functions $J$ \cite{10,11}, all the integrals from the above amplitude can be expressed in the form:
\begin{equation}\label{ws}
\int_0^{\infty} dz zJ_{\mu}(p z)J_{\nu}(p\,' z),
\end{equation}
which are known as Weber-Schafheitlin integrals \cite{10,26}. Some aspects related to these integrals are given in the Appendix.

After we complete the calculations, the final result for the transition amplitude is expressed in terms of Gauss hypergeometric functions $_{2}F_{1}$ , Beta Euler functions $B$ and unit step functions $\theta$ as follows:
\begin{eqnarray}\label{final}
\mathcal{A}_{e^-e^+}=i
\frac{e^{2} Z}{16\pi|\vec{p}+\vec{p}\,'|^{2}}\left[-sgn(\lambda\,')\left(p^{-1}\theta(p-p\,')f_{-k}\left(\frac{p\,'}{p}\right)+
p\,'^{-1}\theta(p\,'-p)f_{k}\left(\frac{p}{p\,'}\right)\right)\right.\nonumber\\
\left.+sgn(\lambda)\left(p^{-1}\theta(p-p\,')f_{k}\left(\frac{p\,'}{p}\right)+
p\,'^{-1}\theta(p\,'-p)f_{-k}\left(\frac{p}{p\,'}\right)\right)\right]\xi^{+}_{\lambda}(\vec{p}\,)\eta_{\lambda'}(\vec{p}\,\,').
\end{eqnarray}
We specify that in the integrals (\ref{in}) for the case $p\,'>p$ (in the case $p\,'<p$ the analysis is similar) we let the momentum $p\,'$ have a small imaginary part $p\,'\rightarrow p\,'+i\epsilon$ which assures the convergence of our integral and makes the unit step function and the $f_{k}\left(\frac{p}{p\,'}\right),\,f_{k}\left(\frac{p\,'}{p}\right)$ functions well defined .
In equation (\ref{final}) the newly introduced functions $f_{k}\left(\frac{p}{p\,'}\right)$, or $f_{k}\left(\frac{p\,'}{p}\right)$, are defined as:
\begin{eqnarray}\label{f}
&&
f_{k}\left(\frac{p}{p\,'}\right)=ie^{-\pi k}\frac{\left(\frac{p}{p\,'}\right)^{1+ik}}{ch^2(\pi k)}\frac{_{2}F_{1}\left(\frac{3}{2},1+i
k;\frac{3}{2}+i k;\left(\frac{p}{p\,'}\right)^{2}-i0\right)}{B\left(-\frac{1}{2},\frac{3}{2}+ik\right)}\nonumber\\
&&+\frac{\left(\frac{p}{p\,'}\right)^{1+i k}}{ch^2(\pi k)}\frac{_{2}F_{1}\left(\frac{3}{2},1+i
k;\frac{3}{2}+ik;\left(\frac{p}{p\,'}\right)^{2}-i0\right)}{B\left(-ik,\frac{3}{2}+ik\right)}-\frac{\left(\frac{p}{p\,'}\right)^{-ik}}{ch^2(\pi k)}\frac{_{2}F_{1}\left(\frac{1}{2},1-i k;\frac{1}{2}-i
k;\left(\frac{p}{p\,'}\right)^{2}-i0\right)}{B\left(ik,\frac{1}{2}-ik\right)}\nonumber\\
&&+ie^{\pi k}\frac{\left(\frac{p}{p\,'}\right)^{-ik}}{ch^2(\pi k)}
\frac{_{2}F_{1}\left(\frac{1}{2},1-i k;\frac{1}{2}-i
k;\left(\frac{p}{p\,'}\right)^{2}-i0\right)}{B\left(\frac{1}{2},\frac{1}{2}-ik\right)}\,\,,
\end{eqnarray}
with the observation that $f_{-k}\left(\frac{p}{p\,'}\right)$ is obtained when one makes the substitution $k\rightarrow-k$ and $f_{k}\left(\frac{p\,'}{p}\right)$ is obtained when one makes the substitution $p\rightleftarrows p\,'$. To simplify our formulas, we can  introduce the following notation $\chi=p/p\,'(=p\,'/p)$ .
A notable observation is that the result can be expressed without letting the momentum have a small imaginary part, in which case the argument $\chi$ in the hypergeometric functions must be considered in the interval $0\leq\chi<1$. This is the domain of convergence of the
hypergeometric functions, because in the limit $\chi\rightarrow1$
the hypergeometric functions become divergent. In this case, one should naturally ask what happens with the result of our integrals with Bessel functions in the limit $\chi=1$. In this limit, the result of the integrals can be expressed in terms of delta Dirac functions $\delta(p-p\,')$, however, when summed, these terms cancel out between them, for each of the two integrals with Hankel functions.

From the final result given in (\ref{final}) and (\ref{f}), it is obvious that the transition amplitude is nonvanishing only for $p\neq p\,'$ and from  here we conclude that the law of conservation for the modulus of the momentum is lost in de Sitter space-time. One can recall that in this geometry the translational invariance with respect to time is lost and from here it is obvious that the laws of conservation for energy and momentum are lost. This means more precisely that the absence of Poincare invariance allows energy and momentum to appear from vacuum \cite{12}. Our result related to the loss of the momentum conservation law, can also be seen as a consequence of the redshifting of the physical momentum in this geometry, $p_{\hat{\mu}}=e^{\nu}_{\hat{\mu}}p_{\nu}$ :
\begin{equation}\label{pm}
p_{ph}=p_{c}/a(t)=p_{c}e^{-\omega t},
\end{equation}
where $p_{c}$ represents a comoving momentum.

It is also more intuitive to imagine the following situation. We know that the energy and momenta are conserved for a particle which interacts with a constant external field. So in our case it is not surprising that the  momentum conservation is broken due to the fact that the perturbation potential acts as an external time dependent source. Thereby, we conclude that the expansion of the spacetime will favor the production of electrons and positrons from vacuum with different modulus of momenta $p\,'\neq p$, in an external Coulomb field.

It is clear that the contribution which is specific to the de Sitter geometry in our amplitude is contained in the function $f_{k}(\chi)$. This quantity encodes via the dependence of the parameter $k=m/\omega$, the effect of the expansion of space on the pair production process and, it is the quantity that enters in our probability and help us to take a closer look at the limit cases.
For studying the validity of our result, in the sense that the functions defined in (\ref{f}) are not divergent, we plot the real part (Re[f]) and imaginary part (Im[f]) of $f_{k}(\chi)$ as function of the parameter $k$ for different values of $\chi=p/p\,'\,\,(p\,'/p)$:

\begin{figure}[h!t]
\includegraphics[scale=0.5]{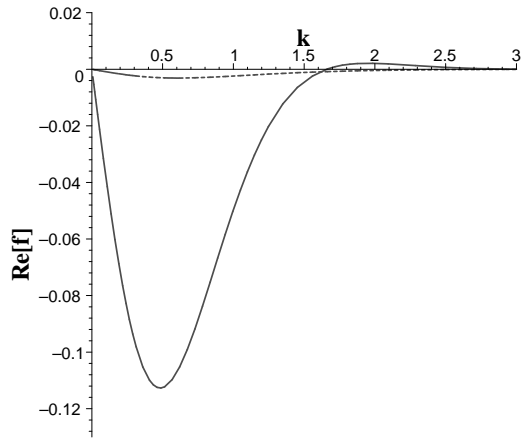}
\caption{The real part of $f_{k}(\chi)$ as a function of $k$. The solid line is for $\chi=0.1$ and the dashed line for $\chi=0.9$.}
\label{f1}
\end{figure}

\newpage
\begin{figure}[h!t]
\includegraphics[scale=0.5]{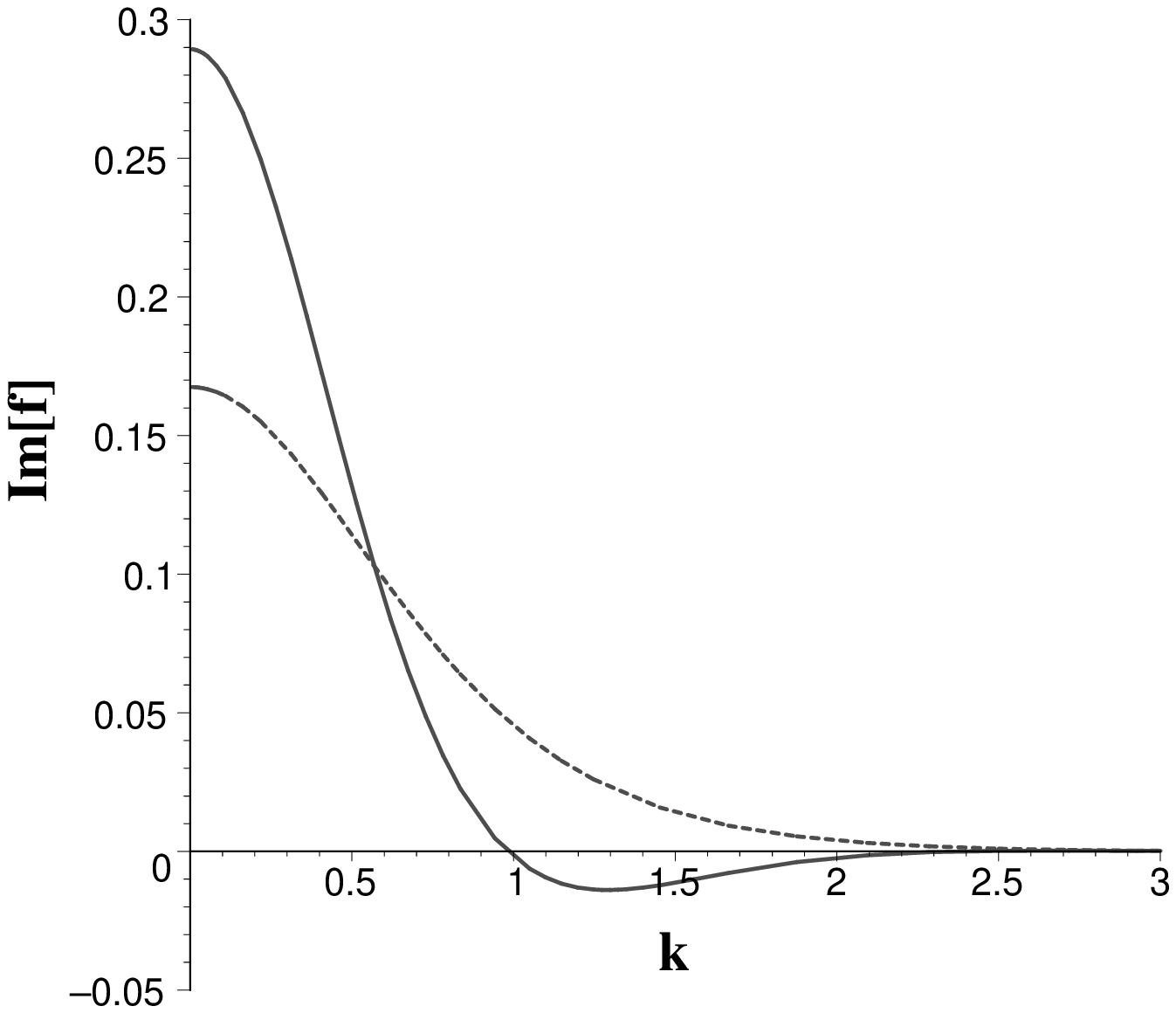}
\caption{The imaginary part of $f_{k}(\chi)$ as a function of $k$. The solid line is for $\chi=0.1$ and the dashed line for $\chi=0.9$.}
\label{f2}
\end{figure}

\begin{figure}[h!t]
\includegraphics[scale=0.5]{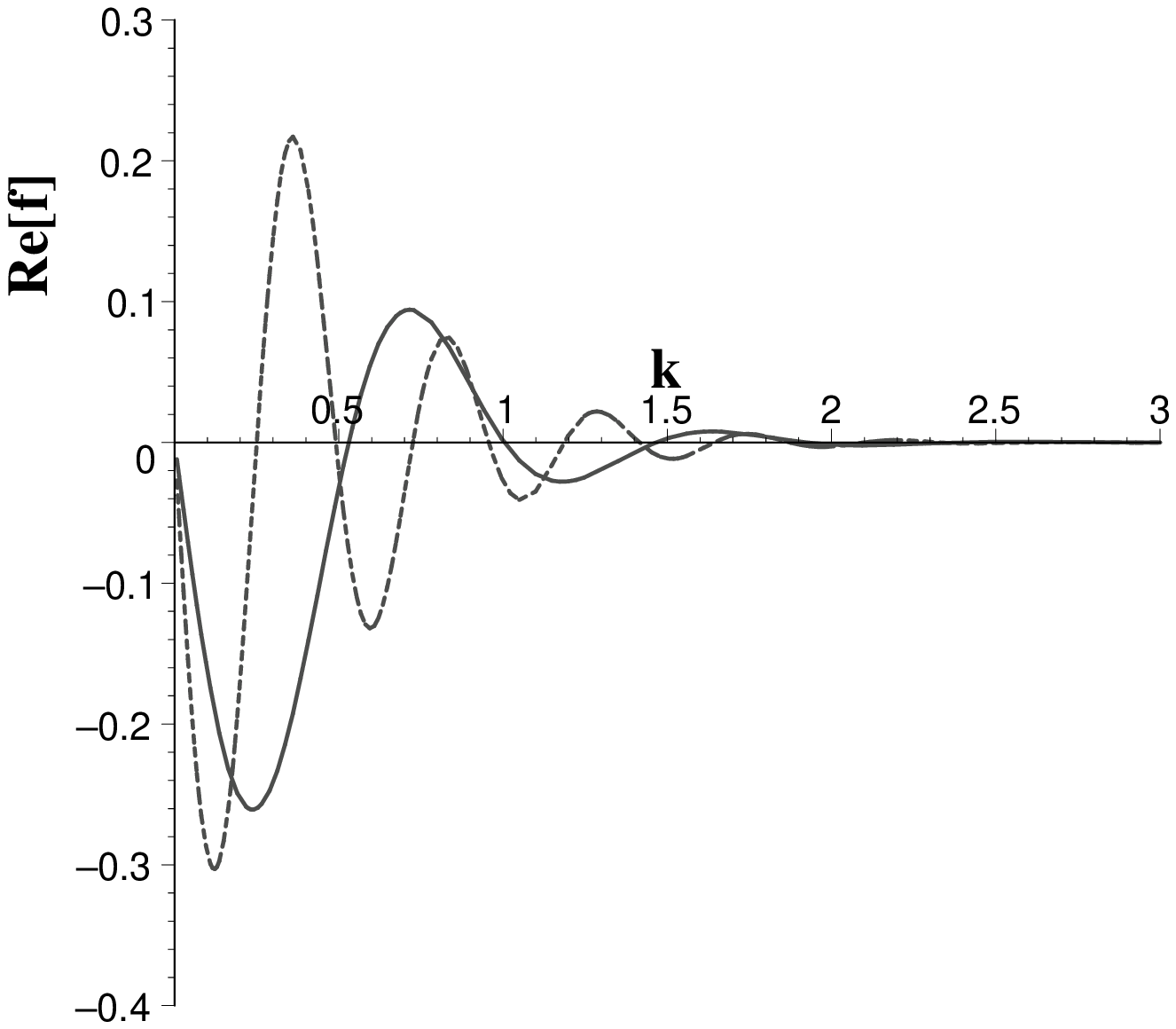}
\caption{The real part of $f_{k}(\chi)$ as a function of $k$. The solid line is for $\chi=0.001$ and the dashed line for $\chi=0.000001$.}
\label{f3}
\end{figure}

\begin{figure}[h!t]
\includegraphics[scale=0.5]{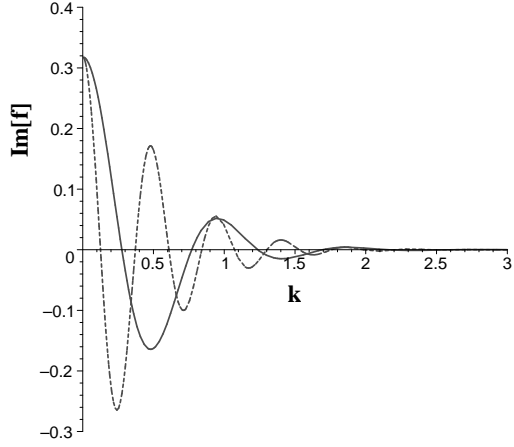}
\caption{The imaginary part of $f_{k}(\chi)$ as a function of $k$. The solid line is for $\chi=0.001$ and the dashed line for $\chi=0.000001$.}
\label{f4}
\end{figure}

\newpage
Our graphs show that both the real part and the imaginary part of the function $f_{k}(\chi)$ are finite in origin and are very  convergent for large values of $k$ (see Figs.(\ref{f1})-(\ref{f4})).
As $\chi$ takes small values, we observe that these functions become oscillatory (see Figs.(\ref{f1})-(\ref{f2})). This oscillatory behavior is the result of the behavior of Gauss hypergeometric functions as their algebraic argument $\chi^2$ approaches zero, combined with the oscillatory factors $\chi\,^{\pm ik}$.

Because we use here the methods based on perturbations, the outcome of our calculations must be the probabilities of transitions. Our results for the amplitude are used in defining the probabilities of fermion production in de Sitter space. The main quantity that enters in the probability is $|\mathcal{A}_{e^-e^+}|\,^{2}$, so we will focus on studying its properties.
We can write down the expression of the probability of transition by squaring the amplitude:
\begin{eqnarray}
&&\mathcal{P}_{e^-e^+}=\frac{1}{2}\sum_{\lambda\lambda'}|\mathcal{A}_{e^-e^+}|\,^{2}\nonumber\\
&&=\frac{1}{2}\sum_{\lambda\lambda'}\frac{e^{4} Z^2}{256\pi^2|\vec{p}+\vec{p}\,'|^{4}}
|\xi^{+}_{\lambda}(\vec{p}\,)\eta_{\lambda'}(\vec{p}\,\,')|^2
\{p\,'^{-2}\theta(p\,'-p)\left[\left|f_{k}\left(\frac{p}{p\,'}\right)\right|^2+\left|f_{-k}\left(\frac{p}{p\,'}\right)\right|^2\right.\nonumber\\
&&\left.-sgn(\lambda)sgn(\lambda\,')\left(f_{k}\left(\frac{p}{p\,'}\right)f^{*}_{-k}\left(\frac{p}{p\,'}\right)+
f^{*}_{k}\left(\frac{p}{p\,'}\right)f_{-k}\left(\frac{p}{p\,'}\right)\right)\right]+
p^{-2}\theta(p-p\,')\left[\left|f_{k}\left(\frac{p\,'}{p}\right)\right|^2\right.\nonumber\\
&&\left.+\left|f_{-k}\left(\frac{p\,'}{p}\right)\right|^2-sgn(\lambda)sgn(\lambda\,')\left(f_{k}\left(\frac{p\,'}{p}\right)
f^{*}_{-k}\left(\frac{p\,'}{p}\right)+
f^{*}_{k}\left(\frac{p\,'}{p}\right)f_{-k}\left(\frac{p\,'}{p}\right)\right)\right]\}.
\nonumber\\\label{prob}
\end{eqnarray}
The total probability of electron-positron pair production in a Coulomb field will be:
\begin{equation}
\mathcal{P}^{tot}_{e^-e^+}=\int \mathcal{P}_{e^-e^+}\,\frac{d^3p}{(2\pi)^{3}}\frac{d^3p\,'}{(2\pi)^{3}},
\end{equation}
and via the squared amplitude $|\mathcal{A}_{e^-e^+}|\,^{2}$ it is also determined by the factor $k=m/\omega$.
Carrying out the integration over $p\,,p\,'$ is a complicated task. The form of the integrals that need to be solved are not known in the literature and only a numerical estimation can be made. However, some physical consequences can be obtained from (\ref{prob}) by drawing the graph of probability of transition $\mathcal{P}_{e^-e^+}$ as a function of $k$, for different values of the ratio $\chi=p/p\,'\,\,(p\,'/p)$.
As we know from Minkowski QED, the helicity conservation is a property which becomes important in the ultra relativistic regime. Let us see what happens in the not so simple de Sitter case. From the above formula for probability of transition, we observe an interesting property, namely that in de Sitter geometry there are nonvanishing probabilities for both production processes which conserve or do not conserve the helicity. In the process under study it is clear that helicity is conserved when $\lambda=-\lambda\,'$, so that the initial zero helicity equals the sum of the final helicities $\lambda+\lambda\,'=0$. Contrary, the process in which helicity is not conserved is obtained when $\lambda=\lambda\,'$. It is interesting to obtain from here some quantitative results from the probability formula (\ref{prob}). These results will help us understand if the expansion of the spacetime will favor pair production processes with conserved/non-conserved helicity.

First we represent our probability of transition (\ref{prob}) as a function of $k=m/\omega$, for $\chi=0.1$ and $\chi=0.4$, both corresponding to close values of the modulus for the two momenta $p\,'\sim p$, but not equal. For $\chi=0.9$, our graphical analysis shows that the probability for a helicity non-conserving process is zero, and for this reason we begin our graphical representation with $\chi=0.4$.
\begin{figure}[h!t]
\includegraphics[scale=0.5]{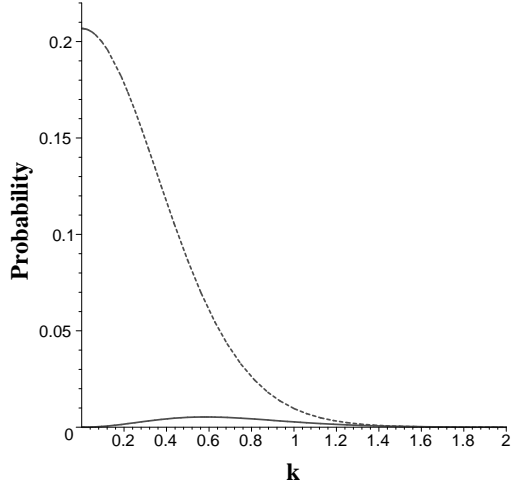}
\caption{$\mathcal{P}_{e^-e^+}$ as a function of $k$ for $\chi=0.4$. The dashed line represents the case of helicity conservation and the solid line the case when helicity is not conserved.}
\label{f5}
\end{figure}

\begin{figure}[h!t]
\includegraphics[scale=0.5]{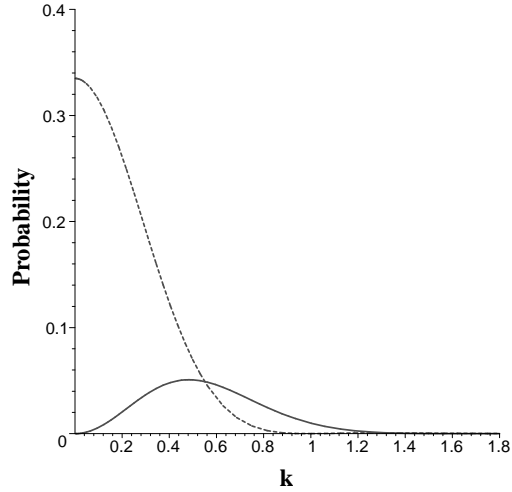}
\caption{$\mathcal{P}_{e^-e^+}$ as a function of $k$ for $\chi=0.1$. The dashed line represents the case of helicity conservation and the solid line represents the case when helicity is not conserved.}
\label{f6}
\end{figure}

\newpage
Our numerical and graphical analysis above show that the probability for pair production with opposite helicities $\lambda=-\lambda'$ (helicity conservation) is sensibly bigger than the probability for production of pair with equal helicities $\lambda=\lambda'$ (helicity non-conserving case) in the case $\omega>m$, (see Figs. (\ref{f5})-(\ref{f6})). For $m=0$, the probability of pair production in the helicity non-conserving case is zero, while the probability for a conserving helicity process is finite. We note that the zero probability of production of fermions with zero mass in the helicity non-conserving case is consistent with the identically null amplitude in the case when  helicity is not conserved due to conformal invariance.  Also, the graphs (see Figs. (\ref{f5})-(\ref{f6})), show that when $m/\omega$ increases the probability of pair production with opposite helicities decreases and the probability of production of fermions with equal helicities becomes finite . Summarizing our result for $\omega>m$, if the two momenta are close as moduli $p\,'\sim p$ but not equal, the expansion of the space will favor production processes which preserve the helicity conservation. This result is in good agrement with  \cite{23}, where the helicity conservation was discussed in the Coulomb scattering on de Sitter space. The result from \cite{23} proves that helicity conservation in Coulomb scattering is preserved as long as we deal with equal initial and final momenta $p_{i}= p_{f}$.

We also see from Figs. (\ref{f5})-(\ref{f6}), that our probabilities in both conserving/non-conserving helicity cases, become negligible for large $m/\omega$, and from here we conclude that the production rate was important only in the inflationary regime, when this ratio was very small . Looking at the helicity non-conserving case ($\lambda=\lambda'$), we obtain from our graphical analysis that for $\chi=0.4$, or equivalently as $\chi$ is closer to unity, the probability for a process where the helicity is not conserved becomes very small.
The conclusion is that in de Sitter space helicity conservation seems to be a property which is preserved as long as we deal with pair production that have the ratio of the momenta $p/p\,'$ closer to unity.

\begin{figure}[h!t]
\includegraphics[scale=0.5]{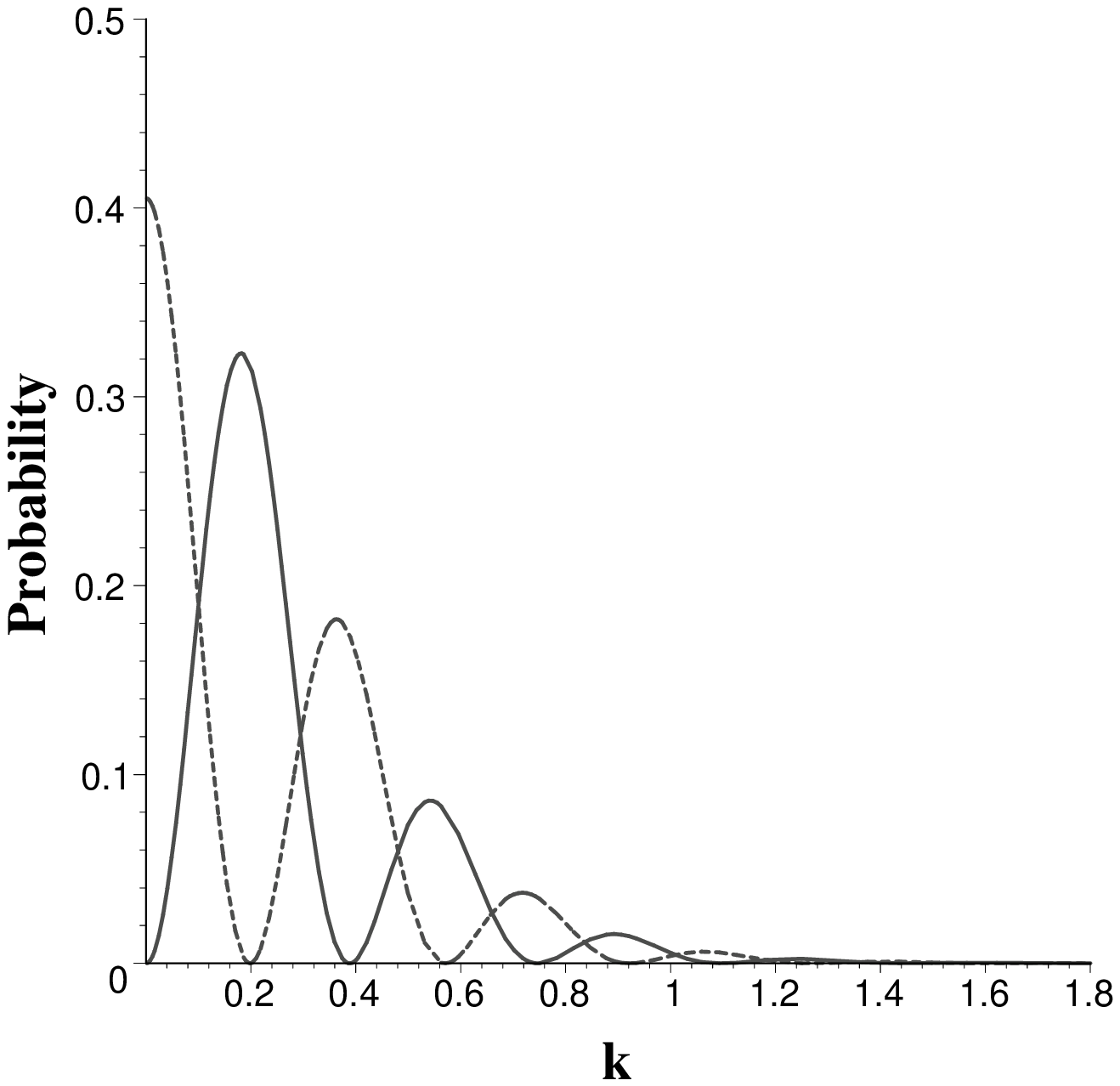}
\caption{$\mathcal{P}_{e^-e^+}$ as a function of $k$ for $\chi=0.0001$. The dashed line represents the case of helicity conservation and the solid line represents the case when helicity is not conserved .}
\label{f8}
\end{figure}

\begin{figure}[h!t]
\includegraphics[scale=0.5]{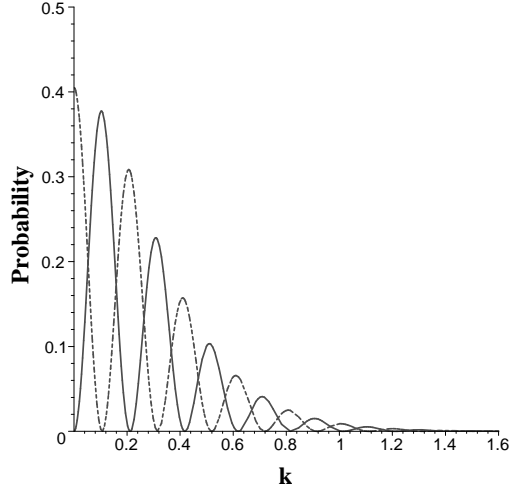}
\caption{$\mathcal{P}_{e^-e^+}$ as a function of $k$ for $\chi=0.0000001$. The dashed line represents the case of helicity conservation and the solid line represents the case when helicity is not conserved .}
\label{f9}
\end{figure}

Let us consider now the case when $\chi$ is very small. This case corresponds to the situation in which, for example, $p\,'\gg p$ (for $p\,' \ll p$ the analysis is similar), so that the momentum of the particle $p$ is much smaller than the momentum of antiparticle $p\,'$. We must specify that large momenta doesn't refer here to a relativistic momentum, but rather to the situation when one of the momenta is almost zero and the other has a value much smaller comparatively with the relativistic case, i.e $p\in(10^{-7},10^{-2})$ and $p\,'\in(1,5)$. Plotting our probability (\ref{prob}) as a function of $m/\omega$ we obtain the results shown in Figs. (\ref{f7})-(\ref{f9}):
\begin{figure}[h!t]
\includegraphics[scale=0.5]{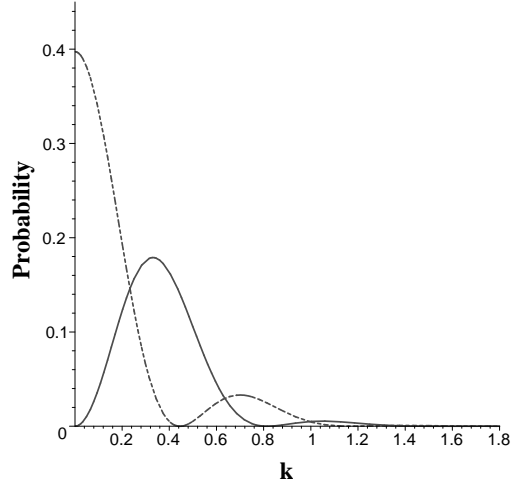}
\caption{$\mathcal{P}_{e^-e^+}$ as a function of $k$ for $\chi=0.01$. The dashed line represents the case of helicity conservation and the solid line represents the case when helicity is not conserved.}
\label{f7}
\end{figure}

\newpage

\newpage
The first observation in this case is that the probability for a production process were helicity is not conserved, becomes important as $\chi\ll 1$, (Fig.(\ref{f9})). We see that in this limit the pair production processes that break the helicity conservation law become important as the two probabilities that correspond to the helicity conservation/non-conservation processes become approximatively equal (Fig.\ref{f9}), for $\omega > m$. Another interesting result is that the probabilities for pair production in both conserving/non-conserving helicity cases have a series of maxima (resonances) and minima (Fig.(\ref{f7})-(\ref{f9})), with the observation that the amplitudes of these resonances decrease as the ratio $m/\omega$ increases. This oscillatory behavior of the probability becomes more pronounced as $\chi$ approaches zero.
The limit $\chi\rightarrow 0$ could help us understand the case in which one of the produced particles has a small momentum close to zero comparatively with momentum of the other particle. Let us make a more detailed analysis of this situation. It is known that in an expanding de Sitter universe the following first order processes could arise as the result of the loss of energy and momentum conservation laws \cite{15,16,17,20,21}: the electron or positron can absorb or can emit photons $(e^{\mp}\leftrightarrows e^{\mp}+\gamma)$, one photon electron-positron pair creation or annihilation of the pair in a single photon $(\gamma\leftrightarrows e^{-}+e^{+})$ and that electron-positron and photon triplet could be produced or absorbed by the de Sitter vacuum, all effects arising as perturbative phenomena. The case $\chi\ll 1$ could be now explained in terms of the above mentioned reactions. One of the Dirac particles arising from vacuum in a Coulomb field could absorb one photon, and if we take in consideration that this photon can be produced from vacuum in the above mentioned reactions, we can resume that the Dirac particle absorbs energy and momenta from vacuum. Opposite to this situation, the emission of one photon is responsible for the small momentum of the other Dirac particle. The emitted photon could end up in another electron-positron pair which could annihilate in vacuum, resulting that the momentum of the particle can be absorbed by the vacuum. Our remarks also help us to understand why the helicity is no longer conserved in the case $\chi\ll 1$. If we start, for example, with an electron, it could emit or absorb one photon and the spin in this reaction is no longer conserved. As a final conclusion, in the limit of small $\chi$ we establish that helicity conservation is no longer a rule and the processes where helicity is not conserved become equal in importance.

It is also important to mention that our graphical results for probability (Figs.(\ref{f5})-(\ref{f9})), are valid for small momenta ($p,\, p\,'$). This is because the factor $|\vec{p}+\vec{p}\,'|^{-4}$ cancel our probability in the limit of large momenta. Thus we can conclude that particles with low momenta were produced in external Coulomb field, due to the early expansion of the space. This is in agreement with the formula for physical momentum (\ref{pm}), from which we see that for large expansion factor $\omega$ the physical momentum become small.

As a general observation from all the graphs of probability as a function of $k=m/\omega$ it is clear that the amount of particle production was important only for small $k$. This corresponds to the early universe, when the expansion was large comparatively with particle mass ($\omega\gg m$). This result is in good agrement with the previously obtained results \cite{7,8,9}, which also predicts that the amount of particle production in the early universe was important.

Taking in consideration that the particle production rate was important for small values of $k\ll1$, it will be interesting to discuss this case. This is the case of a large expansion factor, with non-zero mass. The amplitude and probability formulae in this section are used in our analysis and, in addition, we have at our disposal the graphs for probabilities. We specify that the two cases for conserving/non-conserving helicity are considered. For our considerations it will be sufficient to restrict ourselves, for example, to the case $p<p\,'$. We use helicity bispinors, for $\lambda=\lambda\,'$ and $\lambda=-\lambda\,'$, and restrict the analysis only to the square of our amplitude. In an orthogonal local frame $\{\vec{e}_i\}$ we take the electron and positron momenta in the plane $(1,2)$, denoting
their spherical coordinates as $\vec{p}=(p,\alpha,\beta)$ and
${\vec{p}\,}\,'=(p\,',\gamma,\varphi)$, where $\alpha,\, \gamma\in(0,\pi);\,\beta,\, \varphi\in(0,2\pi)$ . Then for $\beta=\pi\,\,,\varphi=0$ and $p<p\,'$ the final result for the square amplitude is:
\begin{eqnarray}
|\mathcal{A}_{e^-e^+}|\,^2=\frac{e^{4} Z^2}{256\pi^2|\vec{p}+\vec{p}\,'|^{4}}
\frac{1}{p\,'\,^{2}}\,\theta(p\,'-p)\left[\left|f_{k}\left(\frac{p}{p\,'}\right)\right|^2+\left|f_{-k}\left(\frac{p}{p\,'}\right)\right|^2
\right.\nonumber\\
\left.\pm f_{k}\left(\frac{p}{p\,'}\right)f^{*}_{-k}\left(\frac{p}{p\,'}\right)\pm
f^{*}_{k}\left(\frac{p}{p\,'}\right)f_{-k}\left(\frac{p}{p\,'}\right)\right]\times
\left\{
\begin{array}{cll}
\cos^2\left(\frac{\alpha+\gamma}{2}\right)&{\rm for}&\lambda=-\lambda'\\
\sin^2\left(\frac{\alpha+\gamma}{2}\right)&{\rm for}&\lambda=\lambda'
\end{array}\right.\label{ij}
\end{eqnarray}
From this formula, if we set $\alpha=\pi\,, \gamma=0$, corresponding to the case when the momenta of the electron and positron are parallel but move in opposite directions, we obtain zero probability for a helicity conservation process, while the probability for a process where helicity is not conserved, becomes maximum. All the particular cases where the momenta are oriented on different directions can now be obtained from (\ref{ij}). We can conclude that for $k\ll1$, the expansion of the space will favor helicity non-conserving processes in which the electrons and positrons have parallel momenta and move in the opposite directions. This result shows that only in the helicity non-conserving processes the chances of separation  between electrons and positrons in the early universe become important. Thus it seems that for $\alpha=0\,, \gamma=0$ the situation is opposite, conserving processes being dominant, with the specification that this case corresponds to the pair having parallel momenta and moving in the same direction, which increases the probability of annihilation of the pair.

In the next section we shall pay a closer attention to the analysis of this probability in the limit cases of the parameter $k$. It will be interesting to study the probability in the limit of large $k$ to make a connection with the well established results from literature, where the transition from vacuum to non-vacuum states are studied using the Bogoliubov transformations. This is because we expect that the behavior of our amplitude/probability for large $k$ to somehow reproduce up to some factors, the results obtained when one uses the asymptotic solutions of the field equations.

\section{Limit cases}
The de Sitter metric reduces to the Minkowski one when the expansion factor of the space vanishes, $\omega=0$. In this limit the free field equations (Dirac and Maxwell equations) also become the field equations from flat space, so we expect our amplitude to reduce in this limit to the corresponding amplitude from Minkowski space, which is zero. In the flat space this process of pair creation from Coulomb field is forbidden by the laws of conservation for energy and momentum. The Minkowski limit corresponds in our case to $k=\infty$, which is the ratio (mass of particle/expansion factor).

Let us see first what happens for large values of the parameter $k$. In this case one can observe from (\ref{f}) that the second and the third arguments in the hypergeometric functions $_{2}F_{1}(a,b;c;z)$ become approximatively equal $(b\simeq c)$ and we can use $_{2}F_{1}(a,b;b;z)=(1-z)^{-a}$ to simplify these functions. For Beta Euler functions one can apply the Stirling formula to obtain an approximation in the limit of large $k$: $\Gamma(z)=z^{z-1/2}e^{-z}\sqrt{2\pi}(1+O(z^{-1})), |z|\rightarrow\infty$. Finally, one finds for large $k$ (only the terms proportional with $e^{\pi k}$ are considered in the functions $f_{k},f_{-k}$):
\begin{equation}\label{asym}
f_{k}(\chi)\simeq \sqrt{\frac{k}{\pi}}\frac{ie^{-\pi k}\chi^{-ik}}{\,(1-\chi^2)^{1/2}}\,\,\,,
f_{-k}(\chi)\simeq \sqrt{\frac{1}{\pi k}}\frac{ie^{-\pi k}\chi^{1-ik}}{\,2(1-\chi^2)^{3/2}},
\end{equation}
with the specification that the term  $e^{-\pi k}$ originates from $ch^{-2}(\pi k)$.
We see that these functions are highly convergent and vanish as:
\begin{equation}
f_{\pm k}(\chi)\sim e^{-\pi k}
\end{equation}
for $k\rightarrow\infty$.
An important observation is that the factor $e^{-\pi k}$ is used in the literature \cite{12, 18,19,27} for characterizing the vacuum to non-vacuum transitions in this geometry. As noted before, the well established results from the literature use the asymptotic behavior of Hankel functions to establish the positive and negative modes and then with the help of the Bogoliubov coefficient $\beta_{p}$, one can obtain the number density of particles per volume $|\beta_{p}|^2 \frac{d^3p}{(2\pi)^{3}}$. For the above mentioned vacuum non-vacuum transitions \cite{12, 18,19,27}, the Bogoliubov coefficient is proportional to $|\beta_{p}|^2 \sim e^{-2\pi m/\omega}$. This exponential factor is also found in our probability for large $k$. Using equation (\ref{asym}) one can obtain the probability of pair production in a Coulomb field for $k\gg\,1$ (or $m \gg\omega$). The cases $p<p\,'$ and $p>p\,'$ are considered in our analysis, with the observation that we no longer need to use step functions. Using the spherical coordinate of $\vec{p}=(p,\alpha,\pi)$ and
${\vec{p}\,}\,'=(p\,',\gamma,0)$, the final result for the probability (conserving and non-conserving helicity processes are included), in the case $m \gg\omega$, is:
\begin{eqnarray}\label{pa}
\mathcal{P}_{e^-e^+}&=&\frac{e^{4} Z^2\,\,e^{-2\pi \frac{m}{\omega}}}{256\pi^3|\vec{p}+\vec{p}\,\,'|^{4}}\frac{p\, p\,' }{\left(p\,'\,^2-p\,^2\right)^2}\left\{
\begin{array}{cll}
\cos^2\left(\frac{\alpha+\gamma}{2}\right)\,,\,&{\rm for}&\lambda=-\lambda'\\
\sin^2\left(\frac{\alpha+\gamma}{2}\right),\,&{\rm for}&\lambda=\lambda'.
\end{array}\right.
\end{eqnarray}
Recently, in \cite{27}, it was shown that the condition $m \gg\omega$ must be imposed in order to obtain production of a thermal spectrum of radiation, as calculated in the work done by Gibbons and Hawking \cite{25}. The exponential in the probability equation (\ref{pa}) has the form of a Boltzmann factor, that means that the spectrum of particles created in our case for $m \gg\omega$ is thermal, having the Gibbons-Hawking temperature $T=\omega/2\pi$. Also, it is remarkable that our probability is proportional with $e^{-2\pi m/\omega}$, which was found in \cite{12,18,19,27}, to be the factor which characterizes the probability of particle production from vacuum for $m \gg\omega$ .

Now let us make an estimation of this probability for the present day expansion of the Universe. The last astronomical observations give  $\omega=2,3\,\cdot10^{-18}s^{-1}$, for the Hubble constant. Then from the ratio $m c^{2}/\hbar\omega\sim5\cdot10^{37}$, (where $m$ is the electron mass) one can see that the large mass approximation is good in considering the pair production at present. It is immediate from (\ref{pa}), in which we replace the above mentioned ratio, that the probability of pair production at present is very small, and we can say that there is no fermion production of finite mass in Coulomb field due to present day expansion, a result which is in good agrement with the conclusion from \cite{7,8,9}.

We can now address the problem of the Minkowski limit. The graphs (Fig.(\ref{f1})-(\ref{f4})) from the previous section, for functions $f_{k}(\chi)$, invite some comments. First, we see that in the limit $k\rightarrow\infty$, the functions $f_{k}(\chi)$ vanish, which is in agrement with the result obtained analytically, $f_{\infty}(\chi)=0$. From here we recover the Minkowski limit, where particle production in a Coulomb field is forbidden by the laws of conservation for energy and momentum. As $k$ goes to large values the de Sitter amplitude equal the Minkowski one $\mathcal{A}_{e^-e^+}|_{k=\infty}=\mathcal{A}_{e^-e^+}(flat)=0$, which corresponds to the fact that in an asymptotic flat universe these amplitudes vanish. Also it is worth mentioning that for large momenta, $p,p\,'\rightarrow\infty$, the amplitude (\ref{final}) and probability (\ref{prob}) vanishes and the recovery of Minkowski results would probably not be surprising.

Secondly, it is obvious from our graphs (Fig.(\ref{f1})-(\ref{f4})) that our formula for functions  $f_{k}(\chi)$ becomes important when the expansion factor is larger than the mass of the particle. The null mass limit corresponds to the early stage inflation of the universe, when the expansion factor was large comparative to the mass of the particle $\omega\rightarrow\infty$. The $f_{k}(\chi)$ function simplifies in this case (i.e. $_{2}F_{1}(1,1/2;1/2;z)=\,_{2}F_{1}(1,3/2;3/2;z)\simeq(1-z)^{-1}$) and for $k=0$ we obtain:
\begin{equation}\label{fo}
f_{0}(\chi)=\frac{i}{\pi}(1+\chi)^{-1}.
\end{equation}
Also, it is immediate from (\ref{final}) that for $k=0$ the amplitude of transition is nonvanishing only for $\lambda=-\lambda'$. This is in agrement with the observations made in the previous section, that the most probable transitions in the null mase case are those in which the helicity is conserved.
Replacing (\ref{fo}) in the transition amplitude (\ref{final}) and observing that in this case we no longer need to use unit step functions, the final result is:
\begin{equation}\label{foo}
\mathcal{A}_{e^-e^+}|_{(k=0)}=
\frac{e^{2} Z}{4\pi^2|\vec{p}+\vec{p}\,\,'|^{2}}\,\frac{\delta_{\lambda,-\lambda}}{p+p\,'-i0}\,\,\xi^{+}_{\lambda}(\vec{p}\,)\eta_{-\lambda}(\vec{p}\,\,').
\end{equation}
It is remarkable that the amplitude of transition in this limit depends only on the modulus of two momenta $p,\,p\,'$. The same results are obtained from our integrals in (\ref{in}) if we set $k=0$, in which case, the Hankel functions simplify to:
\begin{equation}
H^{(2)}_{1/2}(z)=i\left(\frac{2}{\pi z}\right)^{1/2} e^{-iz},
\end{equation}
and the integrals from (\ref{in}) can be evaluated adding a factor $e^{-\epsilon z}$, where the parameter $\epsilon$ is positive, to assure the convergence in the limit $z\rightarrow\infty$.
Note also that the amplitude in the zero mass limit vanishes if the particle and antiparticle have the same helicities, which corresponds to the non-conserving helicity in the process presented above, which also confirm our graphical results for probabilities. This is an expected result since the Dirac field for a null fermion mass is conformally invariant and the de Sitter metric is conformal with the Minkowski one.

One can also see that for a very large expansion factor but non-zero mass, the exponential factor $e^{-\omega t}$, from the Coulomb potential, implies that only negative large time will be relevant, which means that we can consider $z \simeq e^{-\omega t}$ large everywhere in the integrals (\ref{in}). Translated in our calculations, this implies that we can use the expansion:
\begin{equation}
H^{(2)}_{\mu}(z)=\left(\frac{2}{\pi z}\right)^{1/2} e^{-i(z-\mu\pi/2-\pi/4)},\,\,z\rightarrow\infty
\end{equation}
which differs from $H^{(2)}_{1/2}(z)$ only by the phase factor $e^{i(\mu\pi/2+\pi/4)}$. The conclusion is that for very large expansion factors, the amplitude of this process will be mainly determined by the contributions from the distant past when space was contracted. It is from here immediate that the frequency blueshift in the distant past makes the mass irrelevant, a fact which leads back to a Minkowski conformal theory. 

Further, let us consider a closer look at our amplitude in the null mass case (\ref{foo}). As before, kinetic parameters can be introduced in the orthogonal local frame $\{\vec{e}_i\}$ . We
take the particle and antiparticle momenta in the plane $(1,2)$ denoting
their spherical coordinates by $\vec{p}=(p,\alpha,\beta)$ and
${\vec{p}\,}\,'=(p\,',\gamma,\varphi)$ where $\alpha,\, \gamma\in(0,\pi)$ and $\beta,\,\varphi\in(0,2\pi)$. Then, using the helicity bispinors, one can obtain for $\lambda=-\lambda\,'=\frac{1}{2}$ the final expression of the transition amplitude:
\begin{equation}\label{fooc}
\mathcal{A}_{e^-e^+}|_{(k=0)}=
\frac{e^{2} Z}{4\pi^2|\vec{p}+\vec{p}\,\,'|^{2}}\,\frac{1}{p+p\,'-i0}\,\,
\left[\cos\left(\frac{\alpha}{2}\right)\cos\left(\frac{\gamma}{2}\right)+e^{i(\varphi-\beta)}
\sin\left(\frac{\alpha}{2}\right)\sin\left(\frac{\gamma}{2}\right)\right],
\end{equation}
where the terms in the square brackets were obtained from $\xi^{+}_{1/2}(\vec{p}\,)\eta_{-1/2}(\vec{p}\,\,')$.
From the above formula, one can obtain all the particular cases of interest in which the electron and positron move in the same direction or move in opposite directions. The probability of production of null mass fermions, if we set $\beta=0\,,\varphi=\pi$, will be:
\begin{equation}\label{mo}
\mathcal{P}_{e^-e^+(k=0)}=
\frac{e^{4} Z^2}{32\pi^4|\vec{p}+\vec{p}\,\,'|^{4}}\,\frac{1}{|p+p\,'-i0|^2}\,\,
\cos^2\left(\frac{\alpha+\gamma}{2}\right).
\end{equation}

From the above formula we see that for $\alpha=\pi\,\,,\gamma=0 $ our probability vanishes. This corresponds to the case in which the particle and antiparticle have parallel momenta and move in opposite directions. Now if we set $\alpha=0\,\,,\gamma=0$ the probability takes it's maximum value, and this is the case in which the particle and antiparticle momenta are parallel and they move in the same direction, increasing the probability of annihilation of the pair.
The obvious conclusion is that if the null mass fermions are produced they will rather annihilate each other, than become separate. From here we conclude that there are small chances of production of null mass fermions in the early universe.
As a final remark, one can see that our perturbational calculations in the null mass case is closely to those from the literature \cite{7,8,9,27}, which establishes that there is no production of zero mass fermions in the early universe.

\section{Conclusions}

Fermion pair production in a Coulomb field in a de Sitter expanding universe was considered in our paper. We found the expression of the differential probability for the transition $vacuum \rightarrow e^-e^+$ in the potential $A^{\hat{0}}$. We considered the initial and final states
of the field as exact solutions of the free Dirac equation in de
Sitter space with a defined momentum and helicity. We also found
that the amplitude of pair production depends, in an essential way, on the parameter
$k=m/\omega$. For a vanishing $k$ we recover the expected result
due to the conformal invariance of the theory in the massless
case. The final result proves that the law of conservation for the momentum is lost in de Sitter space as a result of the loss of translational invariance with respect to time. From our amplitude formula, we recover the Minkowski limit where there is no particle production in the Coulomb field. Our final results related to helicities show that the helicity conservation processes seem to be dominant in the case when the momenta of the electron and positron are close in modulus ($\chi\rightarrow 1$). When the ratio of the two momenta is small we found that the processes where helicity is not conserved become important. Our study also led to an important result related to the mechanism of separation between matter and antimatter in the early universe, namely that only in the helicity non-conserving case there seems that are nonvanishing probabilities for producing pairs which move in opposite directions, increasing in this way the possibility of separation between particles and antiparticles.

A notable result is that only particles with small momenta were produced in Coulomb field in the early universe.

Another result is that for the present day expansion, the amount of fermion production in Coulomb field is negligible. Contrary to this, our calculations and graphs shows that in the early stage of the universe the amount of pair production was important. This confirms the results from literature related to the phenomenon of pair production in the early universe.

Our study doesn't take into account the problem of wave packets in the de Sitter geometry. This problem could be important in studying the effects of pair production, and receives little attention in the literature. However, the study from \cite{24} is a first step in this direction, where the construction and propagation of gaussian wave packets on de Sitter geometry was discussed both for scalar and Dirac fields in the null mass case. The study of wave packets in this geometry could improve our understanding of scattering processes and could help us to understand the physical significance of the cross section.

\section{Appendix}
The main steps leading to our amplitude (\ref{final}) will be detailed in this section. First, we will use the well known formula for Hankel functions in terms of Bessel functions $J$ \cite{10,11}:
\begin{eqnarray}
H^{(1)}_{\mu}(z)=\frac{J_{-\mu}(z)-e^{-i\pi\mu}J_{\mu}(z)}{i\sin(\pi\mu)}\nonumber\\
H^{(2)}_{\mu}(z)=\frac{e^{i\pi\mu}J_{\mu}(z)-J_{-\mu}(z)}{i\sin(\pi\mu)}.
\end{eqnarray}

This transformation finally leads to integrals of the type Weber-Schafheitlin \cite{10,26}.
These integrals can be solved in two distinct cases:
\begin{eqnarray}
\int^{\infty}_{0}dzz^{-s}J_{\mu}(\alpha z)J_{\nu}(\beta z)&=&\frac{\alpha^{\mu}\Gamma(\frac{\mu+\nu-s+1}{2})}{2^{s}\beta\,^{\mu-s+1}
\Gamma(\mu+1)\Gamma(\frac{\nu-\mu+s+1}{2})}\nonumber\\
&&\times_{2}F_{1}\left(\frac{\mu+\nu-s+1}{2},\frac{\mu-\nu-s+1}{2};\mu+1;\frac{\alpha^{2}}{\beta^{2}}\right),\nonumber\\
&& \beta > \alpha>0,Re(s)>-1,Re(\mu+\nu-s+1)>0,\label{eq:int1}
\end{eqnarray}
and
\begin{eqnarray}
\int^{\infty}_{0}dzz^{-s}J_{\mu}(\alpha z)J_{\nu}(\beta z)&=&\frac{\beta^{\nu}\Gamma(\frac{\mu+\nu-s+1}{2})}{2^{s}\alpha^{\nu-s+1}
\Gamma(\nu+1)\Gamma(\frac{\mu-\nu+s+1}{2})}\nonumber\\
&&\times_{2}F_{1}\left(\frac{\mu+\nu-s+1}{2},\frac{\nu-\mu-s+1}{2};\nu+1;\frac{\beta^{2}}{\alpha^{2}}\right),
\nonumber\\
&&\alpha > \beta>0,Re(s)>-1,Re(\mu+\nu-s+1)>0,\label{eq:int2}
\end{eqnarray}
both cases being fulfilled in our analysis.

Now one sees that our integrals in (\ref{ws}) demand $s=-1$. This is
problematic because the integral becomes oscillatory for $z
\rightarrow\infty$. Then a simple way to solve this problem is to
consider a parameter $s$ of the form
\begin{equation}
s=-1+\epsilon
\end{equation}
and let, in the end, $\epsilon \rightarrow 0$. No notable
differences appear when evaluating the integrals in (\ref{ws}) directly
with $s=-1$ and with $\mu,\nu=1/2\pm ik$ .
\par
\textbf{Acknowledgements}
\par
I would like to thank Professor Ion I.Cot\u aescu for reading the
manuscript and for suggestions that help me
to improve this work. Also I would like to thank dr.Claudiu Biri\c s for reading the manuscript and to Alexandra Popescu for
helping me with graphical representations.

\end{document}